\documentclass[%
 reprint,
showpacs,preprintnumbers,
natbib,
 amsmath,amssymb,
 aps,
pra,
]{revtex4-1}

\newcommand{\be}{\begin{equation}}
\newcommand{\ee}{\end{equation}}
\newcommand{\bea}{\begin{eqnarray}}
\newcommand{\eea}{\end{eqnarray}}

\usepackage{amsmath}
\usepackage{epsfig}
\usepackage{graphicx}
\usepackage{dcolumn}
\usepackage{bm}
\usepackage{natbib}
\usepackage{cancel}

\begin{document}



\title{Multimode phononic correlations in a nondegenerate parametric amplifier}

\author{S. Chakram, Y.S. Patil and M. Vengalattore}
\affiliation{Laboratory of Atomic and Solid State Physics, Cornell University, Ithaca, NY 14853}
\email{mukundv@cornell.edu}
\begin{abstract}
We describe the realization of multimode phononic correlations that arise from nonlinear interactions in a mechanical nondegenerate parametric amplifier. The nature of these correlations differs qualitatively depending on the strength of the driving field in relation to the threshold for parametric instability. Below this threshold, the correlations are manifest in a combined quadrature of the coupled mechanical modes. In this regime, the system is amenable to back-action evading measurement schemes for the detection of weak forces. Above threshold, the correlations are manifest in the amplitude difference between the two mechanical modes, akin to intensity difference squeezing observed in optical parametric oscillators. We discuss the crossover of correlations between these two regimes and applications of this quantum-compatible mechanical system to nonlinear metrology and out-of-equilibrium dynamics. 
\end{abstract}
\maketitle
\section{Introduction}

The quantum control, detection and manipulation of macroscopic mechanical systems has made enormous strides from its origins in the context of gravitational wave detection \cite{braginsky1980, caves1980, johnson1981} to current efforts on cavity optomechanics \cite{blencowe2004, schwab2005, poot2012, aspelmeyer2013, meystre2013} and quantum-enhanced metrology \cite{hertz2010}. With increasing sophistication of experimental techniques and material platforms amenable to such studies, a broader range of questions have come into focus including the use of such mesoscopic mechanical systems for studies of macroscopic entanglement \cite{armour2002,johansson2014}, out-of-equilibrium thermodynamics \cite{zhang2014,brunelli2014} and the quantum-to-classical transition \cite{ghirardi1986,habib2002,katz2008}. 

A key enabling ingredient for these studies is the realization of mechanical systems with low dissipation and strong, quantum-compatible nonlinear interactions. While mesoscopic mechanical systems exhibit a wide range of mechanical nonlinearities \cite{lifshitz2008}, it is typically the case that such nonlinear effects are weak and only arise at large amplitudes of motion or are present in highly dissipative systems. In either scenario, these preclude quantum-limited operation. Alternately, such nonlinear couplings can also be realized through optical mediation \cite{seok2012, woolley2014}. However, the experimental constraints posed by such optically mediated interactions remain challenging to satisfy. 

An alternate avenue to combining low intrinsic dissipation and strong nonlinear interactions exploits the notion of reservoir engineering \cite{diehl2008,tomadin2012,tan2013,wang2013} - the control of the properties and effective interactions of the system through appropriate design of its environment. Most proposals to date have focused on tuning the properties of an optical reservoir that is coupled to the mechanical system via radiation pressure, a coupling that is typically weak in current optomechanical systems. However, reservoir engineering can also be effected through purely mechanical means, i.e. the interactions between distinct modes of a resonator can be mediated and enhanced by discrete excitations of a massive supporting substrate. Motivated by recent experimental demonstrations of such two-mode parametric nonlinearities in an ultrahigh $Q$ membrane resonator \cite{patil2014}, we describe the various regimes of this system with an emphasis on the nature and fidelity of multimode phononic correlations that arise due to this nonlinear coupling. Due to its compatibility with established techniques of radiation pressure cooling and quantum-limited optical detection, this nonlinear mechanical system offers a powerful platform for studies of nonlinear metrology, the robust generation of entangled mechanical states and the quantum dynamics of mesoscopic mechanical systems.

This paper is organized as follows. In Sec.II we describe the two-mode nonlinearity arising from the parametric interaction between the resonator and its supporting substrate. This interaction realizes a phononic version of a nondegenerate parametric amplifier involving a substrate excitation (pump) and two resonator modes (signal, idler). Such amplifiers are characterized by a threshold pump amplitude beyond which the system is susceptible to self-oscillation. In Sec.III we describe the behavior of this system below threshold. We discuss the onset of two-mode squeezing and calculate the limits of such squeezing in the presence of dissipation, frequency asymmetries and finite pump detuning from parametric resonance. In Sec.IV, we describe the behavior of this system above threshold where the parametric oscillator exhibits amplitude difference squeezing. Finally, in Sec.V we describe the crossover regime around the instability threshold and applications of this system to nonlinear metrology and mesoscopic quantum dynamics. 

\section{Nondegenerate parametric amplifier: Model and Phenomenology\label{sec:2}}

We begin by describing the phenomenology of the two-mode parametric nonlinearity. To motivate the discussion, we consider the physical system described in previous work \cite{patil2014}. The mechanical resonator consists of a silicon nitride (SiN) membrane under high tensile stress that is deposited on a single-crystal silicon substrate. In these membranes, a combination of high intrinsic stress and substrate-induced suppression of anchor loss leads to the robust formation of a large number of mechanical modes with low dissipation and high degree of isolation from the environment \cite{chakram2014}. Furthermore, the supporting substrate can parametrically mediate multimode interactions within the membrane. Such nonlinear interactions are especially significant when the parametric resonance coincides with a discrete excitation of the substrate, as the coupling strength is now enhanced by the quality factor of the mediating excitation (see Fig. 1). 

We describe the parametric two-mode nonlinearity by an interaction of the form $\mathcal{H} = -g X_S x_i x_j$ where $g$ parametrizes the strength of the interaction, $X_S$ is the amplitude of the substrate excitation and $x_{i,j}$ are the amplitudes of the individual membrane resonator modes. 

\begin{figure}[h]
\begin{center}
\includegraphics[width=0.35\textwidth]{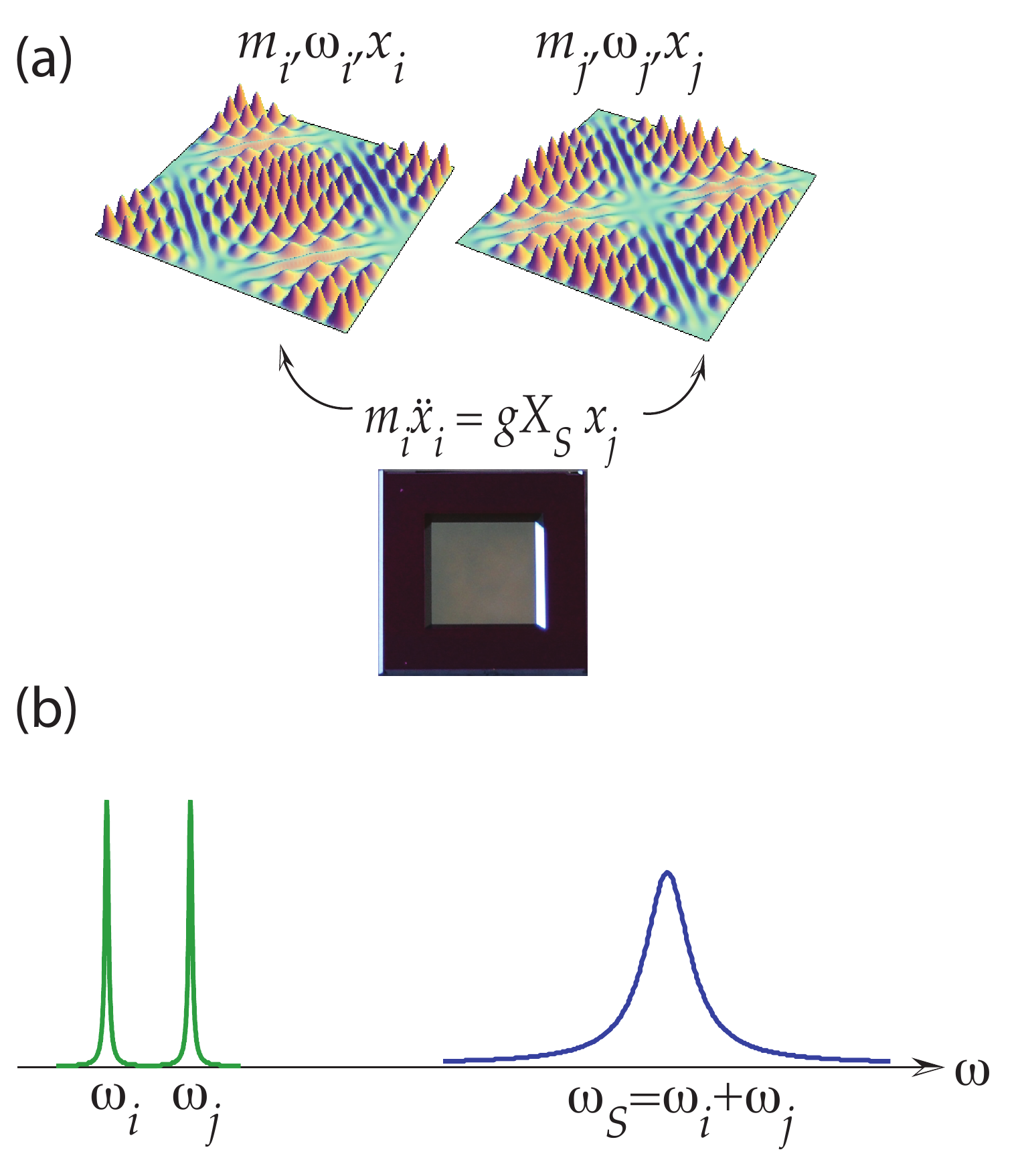}
\caption{(a) Distinct eigenmodes of the membrane resonator are coupled through parametric excitation of the supporting substrate. (b) The strength of the two-mode interaction is enhanced when the parametric interaction between membrane eigenmodes at $\omega_{i,j}$ is mediated by a substrate excitation at $\omega_S = \omega_i + \omega_j$. }
	\label{Fig2_Amplitudes}
	\end{center}
\end{figure} 

Within the rotating wave approximation, this results in equations of motion of the form,
\bea
\ddot{x}_i &+& \gamma_i \dot{x}_i + \omega_i^2 x_i = \frac{1}{m_i} (F_i(t) + \frac{g}{2} X_S x^*_j)\label{fasti}\\
\ddot{X}_S &+& \gamma_S \dot{X}_S + \omega_S^2 X_S = \frac{1}{M_S} (F_S(t) + \frac{g}{2} x_i x_j)\label{fastS}
\eea
with the corresponding equation for mode $j$ obtained by substituting $i \leftrightarrow j$ in Eq.[\ref{fasti}]. Here, we have taken $x_{i,j}, X_S$ to denote the (complex) displacements of the individual modes. The external actuating force and thermomechanical noise forces acting on the various modes are together represented by $F_{i,j,S}$, and $\omega_{i,j,S}$, $\gamma_{i,j,S}$ and  $m_{i,j,S}$ are the eigenfrequencies, mechanical linewidths and the masses of the modes. For now, we assume that $\omega_{S} = \omega_{i} + \omega_{j}$, i.e. the pump actuation is at the parametric resonance.  Lastly, in keeping with the experimental scenario, we assume that the dissipation rate of the substrate excitation is significantly larger than that of the membrane modes ($Q_{S} = \omega_S/\gamma_S \sim 10^{3}-10^4$, $Q_{i,j} = \omega_{i,j}/\gamma_{i,j} \sim 10^{7}$).

The coupled equations of motion can be solved using the methods of two timescale perturbation theory \cite{lifshitz2008}. This gives first order coupled equations of the form,
\bea
2 \dot{A}_{i}  &=& \gamma_{i} \left[ - A_{i}  + i \frac{g}{2}\chi_{i}A^*_{j} A_S +  i\chi_{i}\tilde{F}_{i}(t)\right]\label{slowi} \\
2 \dot{A}_{j}  &=& \gamma_{j} \left[ - A_{j}  + i \frac{g}{2}\chi_{j}A^*_{i} A_S +  i\chi_{j}\tilde{F}_{j}(t) \right]\label{slowj} \\
2 \dot{A}_{S}  &=& \gamma_{S} \left[ - A_{S} + i \frac{g}{2}\chi_{S}A_{i} A_{j} +  i\chi_{S}\tilde{F_{S}}(t)\right]\label{slowS}
\eea
where $x_k = A_k e^{-i \omega_k t},\,\, k\in[i,j,S]$. Also, $\tilde{F}_{k},\,\, k\in[i,j,S]$ are the slowly varying (complex) amplitudes of the external forces on the individual modes, and $\chi_{k} = (m_k \omega_k \gamma_k)^{-1}$ are the magnitudes of the on-resonant susceptibility of the various modes. We have ignored terms such as $\ddot{A}_k, \gamma_i \dot{A}_k$ in the slow time approximation.

As the actuating force on the substrate is increased from zero, the substrate (pump) displacement increases in linear proportion until it reaches a threshold amplitude
\bea
|A_{S,cr}|  = \chi_S \tilde{F}_{S,cr} &=&  \frac{2}{g\sqrt{\chi_{i}\chi_{j}}} \propto {\sqrt{\frac{1}{Q_i Q_j}}}\label{Eq:pump_amp}
\eea 
Below this critical amplitude, the steady-state amplitudes of the membrane modes (signal, idler), denoted by $\bar{A}_{i,j}$, remain at zero and the system realizes a nondegenerate parametric amplifier. Once the parametric drive exceeds the critical value, the membrane modes exhibit an instability to self-oscillation with $\bar{A}_{i,j} \neq 0$, and the system realizes a phononic version of an optical parametric oscillator. At the threshold, the system is characterized by a divergent mechanical susceptibility and response time. The behavior of the parametric system in the vicinity of this threshold can be described in terms of a nonequilibrium continuous phase transition. 

As can be seen from the above expression, the two-mode coupling is enhanced by the quality factors of the individual resonator modes, resulting in strong multimode interaction strengths even in the presence of low dissipation. 

As the pump actuation is further increased, the substrate amplitude remains at the threshold value while the signal and idler amplitudes grow as
\be
|\bar{A}_{i,j}| = \frac{2}{g\sqrt{\chi_{i,j}\chi_{S}}}\sqrt{\mu-1}\label{Eq:probe_amp}
\ee
This behavior of the pump, signal and idler mode amplitudes as a function of the normalized drive, $\mu = \frac{|F_{S}|}{|F_{S,cr}|}$ is shown in Fig.[\ref{Fig2_Amplitudes}].

\begin{figure}[h]
\begin{center}
\includegraphics[width=0.45\textwidth]{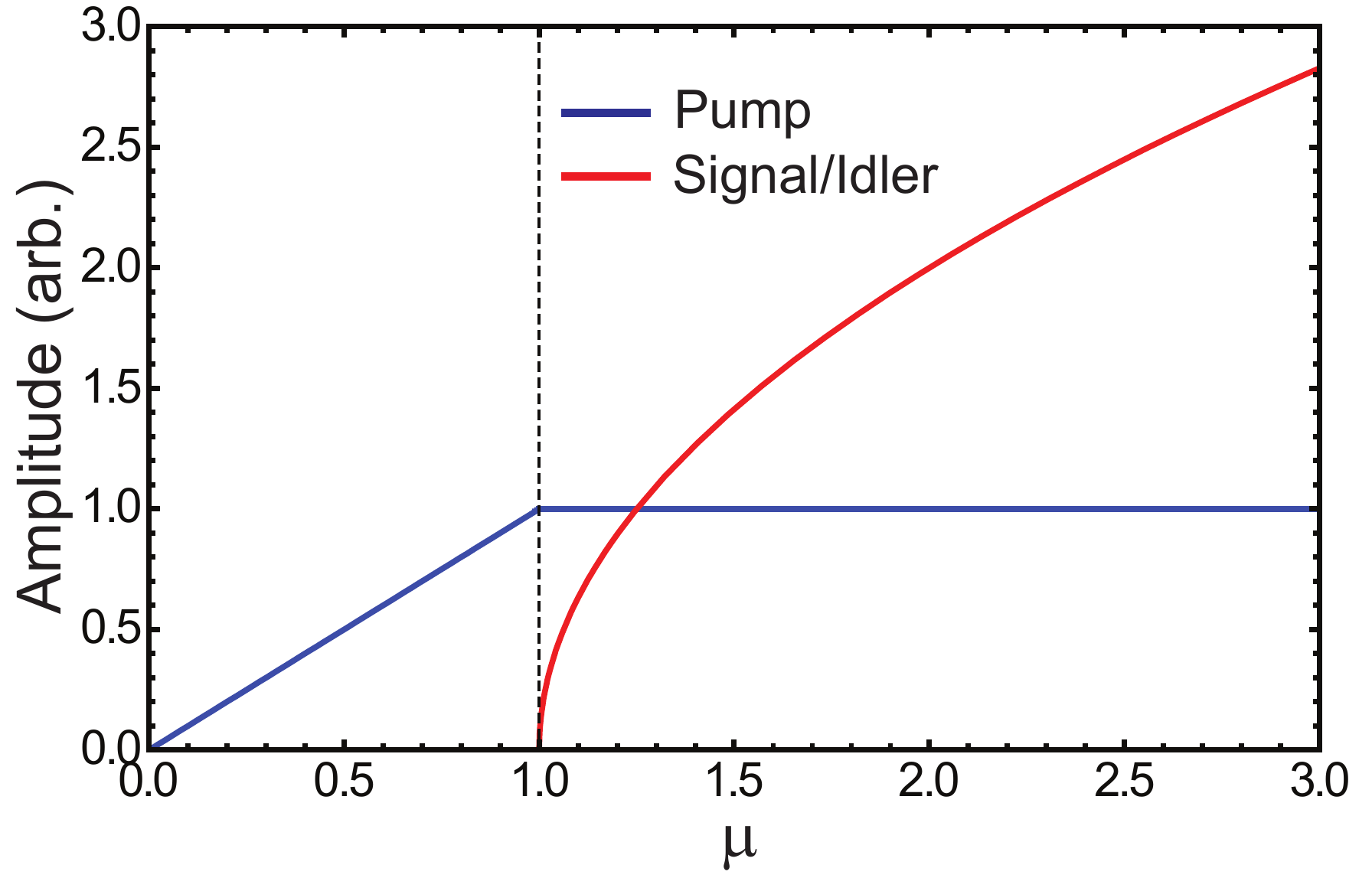}
\caption{Pump (blue) and Signal/Idler (red) amplitudes as a function of the normalized parametric drive, $\mu = \frac{|F_{S}|}{|F_{S,cr}|}$.}
	\label{Fig2_Amplitudes}
	\end{center}
\end{figure} 

This parametric process can be viewed as the down-conversion of one phonon from the pump mode to a pair of phonons, one in each of the signal and idler modes. At the quantum limit, this results in the entanglement of motion of these modes. In the classical regime, this down-conversion is manifest as correlations in the mechanical motion of the two modes. Below threshold, the nonlinear interaction realizes a parametric amplifier with a gain that is dependent on the relation between the phases of the resonator modes and that of the pump. This phase-dependent gain results in thermomechanical squeezing of composite quadratures formed from linear combinations of the quadratures of the individual mechanical modes \cite{patil2014}. 

Above threshold, the rate of phonon down-conversion exceeds the intrinsic loss from either resonator mode, leading to self-oscillation. In this regime, the correlated production of down-converted phonons manifests as a reduction in the variance (squeezing) of the difference in the amplitude fluctuations of the signal and idler modes. This is the thermomechanical analog of intensity difference squeezing in optical parametric oscillators. 

\section{Below threshold dynamics : Two-mode squeezing \label{sec:4}}

In this section, we evaluate the dynamics of the nondegenerate parametric amplifier that arise for pump actuation below the instability threshold. In this regime, the system exhibits correlations between the displacements of the signal and idler modes. These correlations are manifest as two-mode squeezing of a combined quadrature composed from the individual modes. 

In general, the correlations can be obtained using the coupled equations for the membrane modes in the simultaneous presence of a classical pump actuation and thermomechanical Langevin forces. The detailed derivation of the noise spectra is given in Appendix I and we only briefly outline the procedure here. We separate the mean displacement and fluctuations about the mean, by writing $x_{i,j} = (\bar{A}_{i,j} + \delta A_{i,j}) e^{-i \omega_{i,j} t}$ with $\langle \delta A_{i,j} \rangle = 0$. Further, the fluctuations are decomposed into their quadrature components as $\delta A_{i,j} = \delta \alpha_{i,j} + i \delta \beta_{i,j}$. Below threshold, the fluctuations of the individual quadratures are given by Eqs.[38,39] of Appendix I with $\bar{A}_{i,j} = 0$.  

We then define cross-quadratures constructed from $\{\alpha_{i,j},\beta_{i,j}\}$, here normalized to their respective thermomechanical amplitudes, according to the relations,
\bea
x_{\pm} &=& (\alpha_i \pm \alpha_j)/\sqrt{2}\label{xpm}\\
y_{\pm} &=& (\beta_i \pm \beta_j)/\sqrt{2} \label{ypm} 
\eea
The two-mode correlations are manifest as amplification and squeezing of the above quadratures. We represent the fluctuations in these cross-quadratures, along with the fluctuation of the substrate mode in the form of the column vectors $\mathbf{X} = \left(\delta x_{+}, \delta x_{-},\delta x_{S}\right)^{T}$,  $\mathbf{Y} = \left(\delta y_{+},\delta y_{-},\delta y_{S}\right)^{T}$. These are related to the original quadrature fluctuations $\delta\vec{\alpha}$, $\delta\vec{\beta}$ via, 
\bea
\bf{X} = \mathbf{R}\delta\vec{\alpha};\quad \bf{Y} = \mathbf{R}\delta\vec{\beta};\quad \mathbf{R} &=& \frac{1}{\sqrt{2}}\left(\begin{array}{ccc} 1 & 1 & 0  \\ 1 & -1 & 0  \\ 0 & 0 & 1 \end{array}\right)
\eea
Finally, the correlations of the cross quadratures are obtained through a corresponding transformation of the spectral densities ${\bf S}_{\alpha/\beta}(\omega)$, for example,
\be
\mathbf{S}_{X}(\omega) = \left\langle \mathbf{X}(\omega)\mathbf{X}(\omega)^{\dagger} \right\rangle = \mathbf{R}\mathbf{S_{\alpha}}\mathbf{R}^{T} \label{Eq:Sx}
\ee
along with the analogous equation for $\mathbf{Y}$ ($X \rightarrow Y\quad\&\quad\alpha\rightarrow\beta$ in the above). As shown in Appendix I, the degree of squeezing is obtained through the variances of these cross quadratures by integrating the spectra in Eq.[\ref{Eq:Sx}] over the measurement bandwidth.

While this outlines the basis of the calculation, the appearance of such two-mode correlations can be intuitively seen as the result of a coherent interference between the response of the individual resonator modes due to thermomechanical noise and the down-converted field arising from the two-mode nonlinearity. This interference results in a reduction of thermomechanical motion along one quadrature at the expense of amplified motion in the orthogonal quadrature. 

In general, the degree of two-mode noise squeezing is sensitive to various experimental considerations such as the `loss asymmetry' arising from mismatched dissipation rates of the individual membrane modes, their frequency difference and the detuning of the parametric drive from the two-mode resonance. Below, we evaluate the effect of these considerations on the degree of squeezing and find the robust formation of two-mode squeezed states for a wide range of experimental parameters. 

\subsection{Matched losses and frequencies}
For notational convenience, we introduce the loss asymmetry parameter $\delta_\gamma = (\gamma_i - \gamma_j)/(\gamma_i + \gamma_j)$ and frequency mismatch parameter $\delta_{\omega} = (\omega_{i}-\omega_{j})/(\omega_{i}+\omega_{j})$. 
To build an intuition for quadrature squeezing below threshold, we first consider the simplest case of distinct resonator modes with identical frequencies ($\omega_i = \omega_j = \omega$ or $\delta_\omega = 0$) and  identical dissipation rates ($\gamma_i = \gamma_j = \gamma$ or $\delta_\gamma = 0$). 

For this case, the evolution matrices in Eq.[\ref{Mgen}] of Appendix I reduce to,
\begin{equation}
{\bf M}_{\alpha/\beta} = \frac{1}{2} \left( \begin{array}{ccc} -\gamma & \mp\gamma\mu & 0 \\
\mp \gamma\mu & -\gamma &  0 \\
0 & 0 & -\gamma_S \end{array}\right)
\end{equation}

\begin{figure}[h]
\begin{center}
\includegraphics[width=0.45\textwidth]{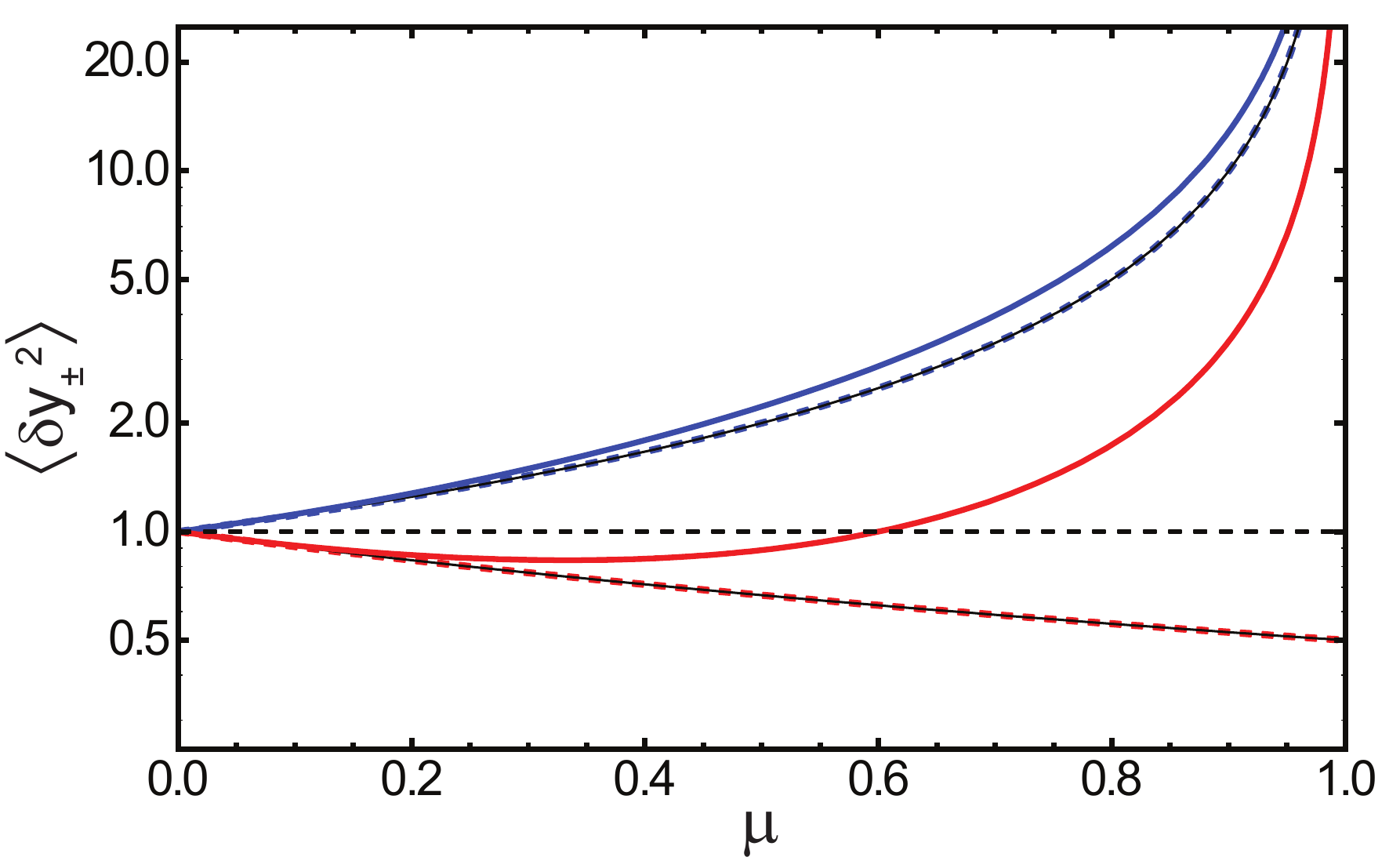}
\caption{Normalized variances of amplified and squeezed collective quadratures as a function of the normalized parametric drive. Black lines indicate the variances for matched frequencies ($\delta_\omega = 0$ ) and loss rates ($\delta_\gamma = 0$). Solid lines indicate the amplified (blue) and squeezed (red) variances for $\delta_\gamma = 0.5$, $\delta_{\omega} = -0.5$. The dashed lines indicate the amplified and squeezed variances for $\delta_\gamma = \delta_\omega \neq 0$, the case of matched asymmetries where the peak noise squeezing again approaches a factor of $\frac{1}{2}$ as $\mu \rightarrow 1$. The dashed horizontal line represents the thermomechanical variance given by $\frac{k_B T}{m \omega^2}$.}
	\label{Fig1}
	\end{center}
\end{figure}

The spectral density of fluctuations of the collective quadratures is evaluated from Eq.[\ref{Eq:spectrum}] and Eq.[\ref{Eq:Sx}] to yield
\be
\mathbf{S_{X/Y}} =  \frac{2}{\pi} \left( \begin{array}{ccc} \frac{\gamma }{\left(\gamma ^2 (1\pm\mu )^2+4 \omega ^2\right)} & 0 & 0 \\
0 & \frac{\gamma}{\left(\gamma ^2 (1\mp\mu )^2+4 \omega ^2\right)} &  0 \\
0 & 0 & \frac{\gamma_{S}}{\left(\gamma_{S} ^2 +4 \omega ^2\right)}  \end{array}\right) \label{Eq:specXY}
\ee
The variances of the normalized collective quadratures of the mechanical modes are then given by,
\be
\sigma_{x_{\pm},x_{\pm}} = \frac{1}{1\pm\mu} = \sigma_{y_{\mp},y_{\mp}}\label{Eq:var}
\ee
We see that $x_{-},y_{+}$ are amplified quadratures with variances that grow as $\mu\rightarrow1$, while $x_{+},y_{-}$ are squeezed quadratures showing reduction in the variance below the thermomechanical limit. For pump actuation close to parametric threshold ($\mu \rightarrow 1$), we obtain a peak noise squeezing of $\frac{1}{2}$, similar to the bound observed in optical parametric amplifiers \cite{milburn1981}. The degree of two-mode squeezing versus parametric drive is shown in Fig.[\ref{Fig1}].

The existence of a bound for the peak noise squeezing can be intuited by looking at the equations of motion for the collective quadratures, through a rotation of the original equations of motion, i.e.
\begin{eqnarray}
\dot{\bf X} &=& {\bf M}_X{\bf X} + {\bf v'}_X \\
\dot{\bf Y} &=& {\bf M}_Y{\bf Y} + {\bf v'}_Y 
\eea
where,
\bea
 {\bf M}_{X/Y}&=&\mathbf{R}\mathbf{M}_{\alpha/\beta}\mathbf{R}^{T} = -\frac{1}{2}\left(\begin{array}{ccc} \gamma(1\pm\mu) & 0 & 0  \\ 0 & \gamma(1\mp\mu) & 0  \\ 0 & 0 & \gamma_{s}\end{array}\right)\nonumber
\end{eqnarray}
and ${\bf v'} = {\bf Rv}$ represents thermomechanical noise forces.  

Two-mode squeezing arises from the fact that while the thermomechanical noise forces remain the same in the presence or absence of the parametric drive, the dissipation of the squeezed quadrature goes to twice its bare value near parametric threshold. This results in a reduction in the variance of the squeezed quadrature by a factor of $2$. Simultaneously, the decay rate of the amplified quadrature goes to zero, signaling the onset of the parametric instability. The onset of the instability thus explains the $3$ dB bound for squeezing through this parametric process. 

The variances of the cross-quadratures are calculated by integrating the noise over all frequencies, i.e. they are measured with infinite bandwidth. As the measurement bandwidth decreases, the peak noise squeezing approaches $6$ dB, i.e.
\be
\frac{S_{x_+,x_+}(\omega=0,\mu)}{S_{x_+,x_+}(\omega=0,\mu=0)} = \frac{1}{(1+\mu)^2} \xrightarrow{\mu \to 1} \frac{1}{4}
\ee

\subsection{Effect of mismatched frequencies and loss rates}
We now consider the case where the frequencies and the damping rate of the two resonators are not the same, i.e. $\delta_\gamma, \delta_\omega \neq 0$. In  this situation, the collective quadratures $x_{\pm}$ and $y_{\pm}$ are no longer decoupled from each other. This coupling between the collective quadratures has been noted to be detrimental to entanglement\cite{szork2014}, and backaction evasion\cite{woolley2014} protocols. It also results in a degradation of peak two-mode thermomechanical squeezing. This is a result of the coupled quadratures $x_{+},x_{-}$ and $y_{-},y_{+}$ being respectively amplified and squeezed in the presence of the parametric drive. 

As before, the variances of the collective quadratures are obtained using Eq.[\ref{Eq:spectrum}], subsequent rotation using Eq.[\ref{Eq:Sx}] and integration over frequency. These are now given by
\bea
\sigma_{y_{\pm},y_{\pm}}&=& \sigma_{x_{\mp},x_{\mp}} \label{Eq:varass}\\ 
&=& \frac{1}{1-\mu^2}\left\{1+\mu^2\frac{\delta_\omega(\delta_\omega-\delta_\gamma)}{1-\delta_\omega^2}\pm\mu\sqrt{\frac{1-\delta_\gamma^2}{1-\delta_\omega^2}}\right\}\nonumber
\eea
with the cross correlation between ($x_{+}, x_{-}$),  ($y_{+}, y_{-}$) given by,
\bea
\sigma_{y_{+},y_{-}}=\sigma_{x_{+},x_{-}}=\frac{\mu ^2 \left(\delta _{\omega }-\delta_\gamma\right)}{2\left(1-\mu ^2\right) \left(1-\delta _{\omega }^2\right)}\label{cross}
\eea
The variances of the amplified and squeezed collective quadratures for the case where ($\delta_\gamma \neq 0, \delta_{\omega} \neq 0$, $\delta_\gamma \neq \delta_\omega$), are shown in Fig.[\ref{Fig1}]. As can be seen, the presence of loss asymmetry or a frequency mismatch results in a degradation of noise squeezing. In this case, optimal squeezing obtained for a parametric drive that is significantly below the instability threshold. We also find that the coupling between the amplified and squeezed quadratures leads to a divergence of the squeezed quadrature at the instability threshold. 

\begin{figure}[h]
\begin{center}
\includegraphics[width=0.5\textwidth]{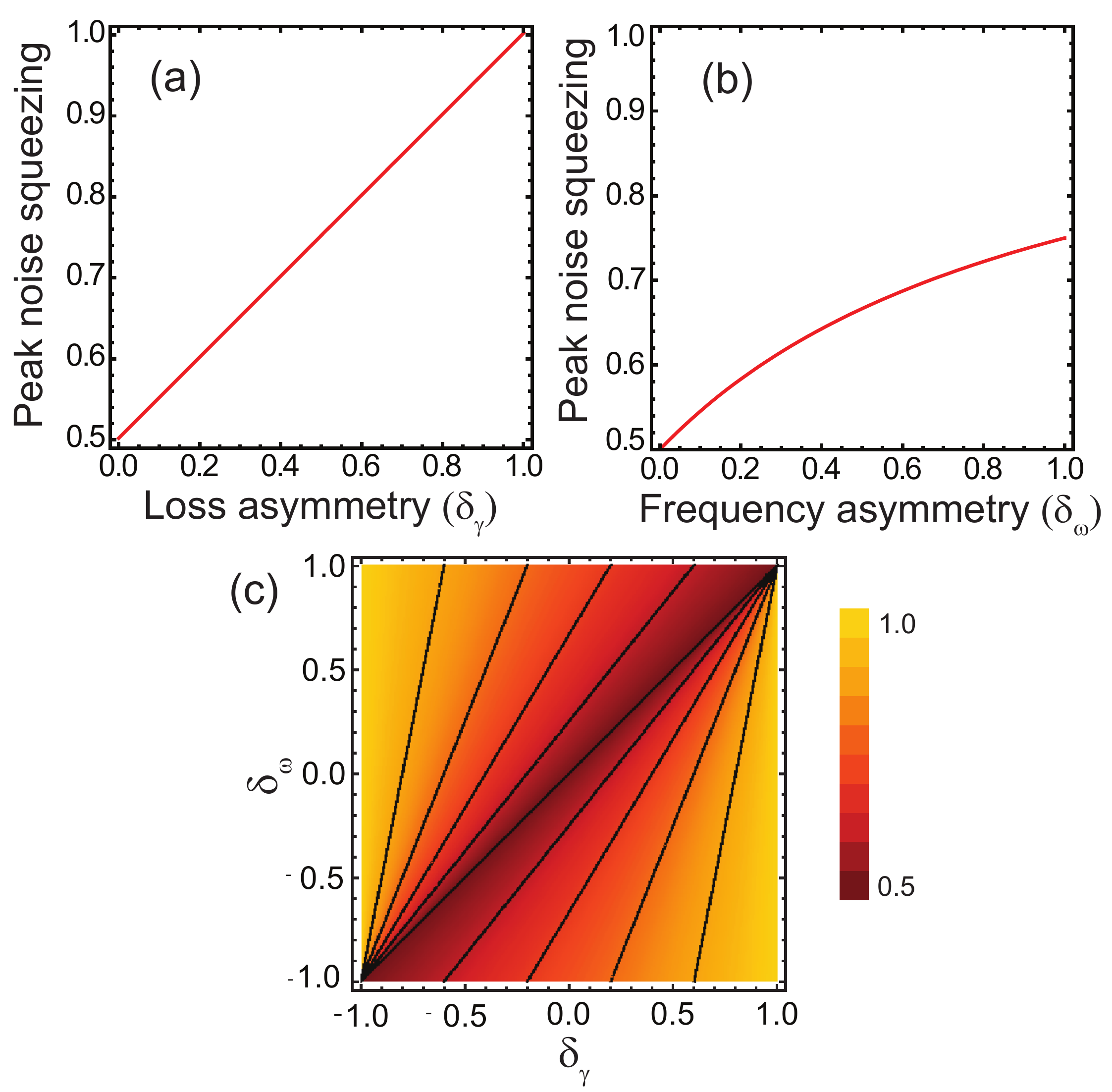}
\caption{(a) Peak noise squeezing as a function of loss assymetry $\delta_\gamma$ for $\delta_\omega = 0$ with the squeezing going linearly from $\frac{1}{2} \rightarrow 1$ as $\delta_\gamma = 0\rightarrow 1$. (b) Peak squeezing as a function of the frequency asymmetry $\delta_{\omega}$, for $\delta_\gamma = 0$. (c) Peak squeezing as a function of $\delta_\gamma$ and $\delta_{\omega}$.}
	\label{Fig2}
	\end{center}
\end{figure}

The dependence of the peak squeezing on the loss asymmetry and frequency mismatch parameters are summarized in Fig.[\ref{Fig2}]. Fig.[\ref{Fig2}(a)] shows a plot of the peak squeezing as a function of the loss asymmetry for the case of distinct mechanical modes with the same frequency ($\delta_\omega = 0$), showing a linear degradation of peak squeezing with loss asymmetry. Correspondingly, Fig.[\ref{Fig2}(b)] shows the peak squeezing as a function of the frequency mismatch parameter ($\delta_\omega$) for the case of no loss asymmetry ($\delta_\gamma = 0$). 

Importantly, as can be seen in Fig.[\ref{Fig2}(c)], we find that the 3 dB squeezing bound can be regained even in the presence of loss asymmetry as long as $\delta_\gamma = \delta_{\omega}$. In this case, the normalized cross correlations between amplified and squeezed collective quadratures vanish, and Eq.[\ref{Eq:varass}] reduces to Eq.[\ref{Eq:var}]. This results in a noise squeezing  that is identical to that for the case of symmetric losses and frequencies.

\subsection{Effect of pump detuning}
Finally, we consider the thermomechanical squeezing bound when the pump drive is detuned from parametric resonance. In this case, the drive frequency is given by $\omega_{d} = \omega_{S} +\Delta$, where $\Delta$ is the drive detuning.  This case is of interest since a detuned parametric drive introduces a dynamic coupling between the $\delta\vec{\alpha}$ and $\delta\vec{\beta}$ quadratures. In turn, this leads to correlations between the amplified and squeezed quadratures. Due to these correlations, the amplified quadrature contains information about the squeezed quadrature that can be used for enhanced localization through weak measurements and optimal estimation \cite{szork2013}. Thus, a detuned parametric drive can allow for enhanced noise squeezing in the presence of feedback. An additional point of interest in this case is that special choices of the drive detuning lead to some of the collective quadratures becoming quantum non-demolition observables\cite{szork2014}. 

 The equations satisfied by the slowly varying complex amplitudes ($A_{i,j,S}$) are again given by Eqs.[\ref{slowi}-\ref{slowS}], with the only difference now being that the pump actuation $\tilde{F}_{S}(t)$ is a slowly varying function of time,  $\tilde{F}_{S}(t) = |\tilde{F}_{S}|e^{-i\Delta t}$. The pump amplitude resulting from this drive force, $\bar{A}_{S}(t)$ is given by,  
\be
\bar{A}_{S} = i\chi_{S}\tilde{F}_{S}(t) = i\chi_{S}|\tilde{F}_{S}|e^{-i\Delta t} = i|\bar{A}_{S}|e^{-i\Delta t}
\ee
Here, as we are most interested in pump detunings that are comparable to the resonator linewidths, we have made the assumption that $\Delta \ll \gamma_S$ and that the pump amplitude is hence related to the instantaneous parametric actuation through the on-resonant susceptibility.

The two-mode correlations are computed in Appendix II, and are given for $\delta_\gamma = \delta_\omega = 0$ by 
\bea
\sigma_{x_{\pm},x_{\pm}} &=& \left(\frac{k_{B}T\gamma}{m\omega^{2}}\right)\Bigg[\frac{\left(\Delta ^2+\gamma ^2 (1\mp\mu)^2-\lambda_{-}^2\right)}{\lambda_{+}\lambda_{-}\left(\lambda_{+}+\lambda_{-}\right)} + \frac{1}{\lambda_{+}}\Bigg]\nonumber\\
&=& \sigma_{y_{\mp},y_{\mp}}
\eea
and
\bea
\sigma_{x_{\pm},y_{\pm}} &=& \left(\frac{k_{B}T\gamma}{m\omega^{2}}\right) \frac{2\Delta\gamma\mu}{\lambda_{+}\lambda_{-}\left(\lambda_{+}+\lambda_{-}\right)} = \sigma_{y_{\pm},x_{\pm}}
\eea
where $\lambda^{2}_{\pm} =\gamma ^2\left(1+\mu ^2\right)-\Delta ^2 \pm 2\gamma\sqrt{\gamma ^2\mu ^2-\Delta ^2}$.

Note that $x_{+}$ and $y_{-}$ are amplified quadratures while $x_{-}$ and $y_{+}$ are squeezed. As mentioned earlier, a nonzero detuning introduces correlations between ($x_{+}, y_{+}$) and ($x_{-},y_{-}$), i.e. between amplified and squeezed quadratures. This is distinct from the case of loss asymmetry or frequency mismatch considered previously, where correlations were introduced between  ($x_{+}, x_{-}$) and ($y_{+},y_{-}$).  As can be seen from the above expressions, these correlations between the $x_{\pm}$ and $y_\pm$ quadratures are proportional to the drive detuning.

\begin{figure}[h]
\begin{center}
\includegraphics[width=0.5\textwidth]{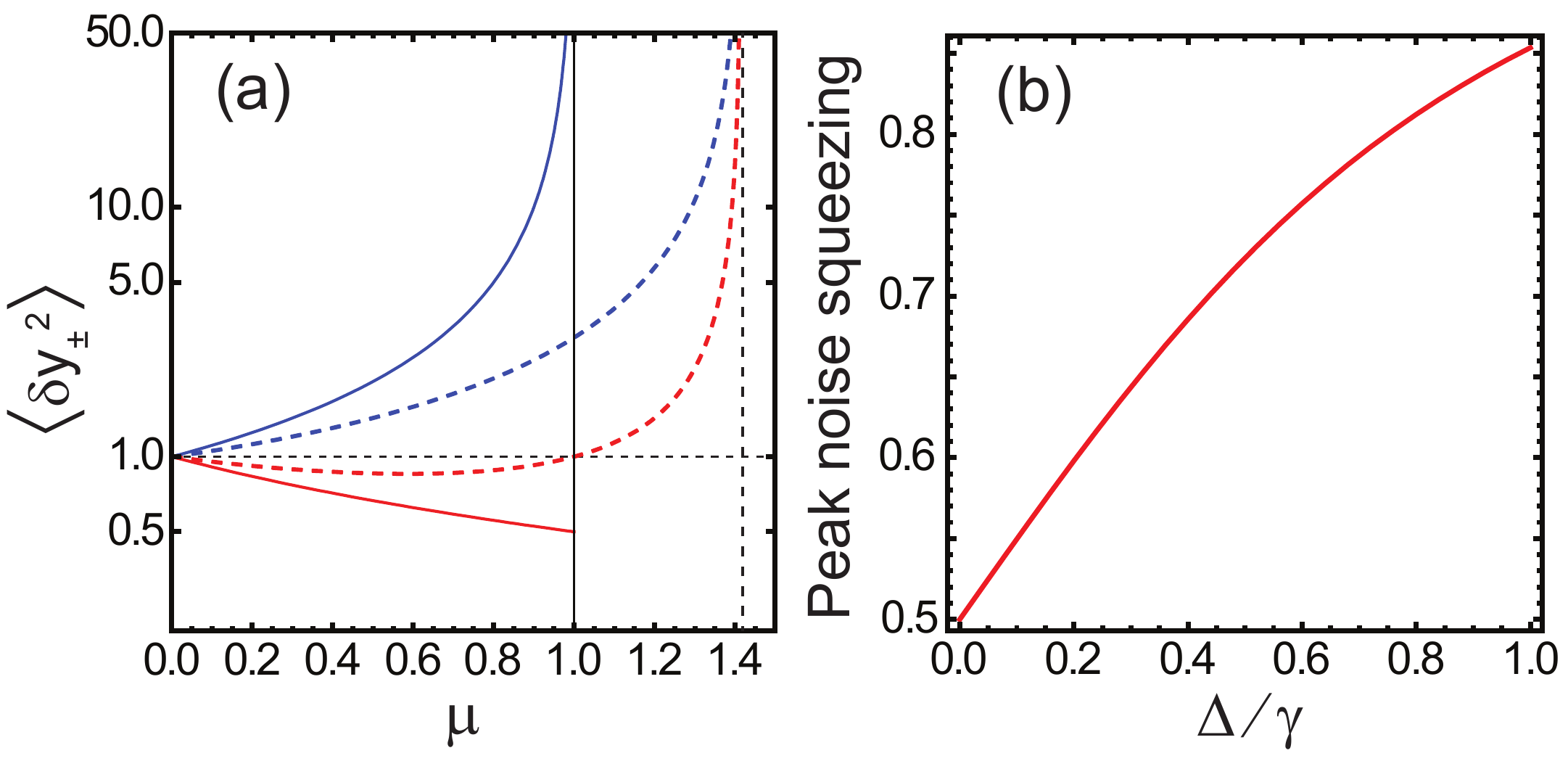}
\caption{(a) Normalized variance of the fluctuations in the $y_{+}$ (amplified,blue) and $y_{-}$ (squeezed,red) quadratures for the case of zero detuning (solid lines) and a detuning of $\Delta = \gamma$ (dashed lines). The  amplified quadrature diverges at the instability threshold. For $\Delta = 0$, this occurs at $\mu = 1$ (solid black vertical line). For $\Delta = \gamma$, this occurs at $\mu = \sqrt{1+(\Delta/\gamma)^{2}}=\sqrt{2}$ (dashed black vertical line). The black horizontal line is at $1/2$. (b) Peak noise squeezing as a function of normalized detuning $\frac{\Delta}{\gamma}$. Both these graphs are shown for the case of no loss or frequency asymmetry, i.e. $\delta_\gamma = \delta_\omega = 0$.}
	\label{Det_Fig1}
	\end{center}
\end{figure}

This coupling between amplified and squeezed quadratures also results in a decrease in the peak squeezing at non-zero detunings, as can be seen in Fig.[\ref{Det_Fig1}(a)]. 
As expected, for the detuned case, the amplified quadrature diverges at $\mu = \sqrt{1+(\Delta/\gamma)^{2}}$, the instability threshold for $\Delta \neq 0$. 

The peak squeezing as a function of the detuning, normalized with respect to the decay rate (for no loss asymmetry or frequency mismatch) is shown in Fig.[\ref{Det_Fig1}(b)]. We see that squeezing is almost completely lost when the detuning becomes comparable to the linewidth of the signal and idler modes.

\section{Above threshold dynamics : Amplitude difference squeezing \label{sec:5}}
For pump actuation above the parametric instability threshold, the two-mode nonlinearity results in parametric self-oscillation of the individual membrane modes. In this regime, the correlated production of down-converted phonons in the signal and idler modes results in a reduction of fluctuations in the amplitude difference between these two modes. This is the phononic version of intensity difference squeezing observed in optical parametric oscillators \cite{fabre1989}. As in the optical case, this form of difference squeezing is of interest in nonlinear interferometric schemes capable of `super-Heisenberg' sensitivities \cite{woolley2008}.

In this section, we discuss the dynamics of the two-mode nonlinearity above threshold and compute the amplitude fluctuations of the mechanical modes. 
Additionally, we discuss the fluctuations of their phases. While energy conservation dictates that the sum of the frequencies of the signal and idler modes equal that of the pump drive, it is also the case that the sum of the phases of the signal and idler modes above threshold remain locked to the phase of the pump drive. This follows from  Eq.[\ref{slowS}], which in steady state can be rewritten in terms of the normalized drive $\mu$ and the drive force phase $\phi_{S}$  ($F_{S} = |F_{S}|e^{i\phi_{S}}$) as
\bea
-(\mu-1)e^{i\phi_{S}} &=& \frac{g}{2} \frac{\chi_{S}}{|A_{S,cr}|} A_{i}A_{j}\label{Pumpph2}
\eea
Defining the phases of the resonator modes as $A_{i,j} = i|A_{i,j}|e^{i\phi_{i,j}}$, we see that $\phi_{i} +\phi_{j} = \phi_{S}$. 
However, the difference in the phases is not constrained and is free to fluctuate. 
 
As before, we quantify these fluctuations in the amplitudes and phases about their steady state values by decomposing the complex amplitude fluctuations $\delta\vec{A}$ into $\delta\alpha_{i}$ and $\delta\beta_{i}$ quadratures. The equations of motion of $ \delta\vec{\alpha}$ and $ \delta\vec{\beta}$ are given by Eqs.[38,39] in Appendix I after substituting $|\bar{A}_{S}|=\sqrt{\frac{\gamma_{i}\gamma_{j}}{\kappa_{i}\kappa_{j}}}$ and $\bar{A}_{i,j}=\sqrt{\frac{\gamma_{i,j}\gamma_{S}}{\kappa_{i,j}\kappa_{S}}}\sqrt{\mu-1}$. Here, we have defined the coupling parameter $\kappa_k = \frac{g}{2 m_k \omega_k}$. 

We choose the drive and resonator mode phases such that $\bar{\phi}_{i,j} = \phi_{S} = 0$. Given this choice of phases, the complex mean amplitudes $\bar{A}_{i,j}$ and the fluctuations are as shown in the inset of Fig.[\ref{Fig4}]. With this convention, the amplitude fluctuations are given by $\delta \beta_{i,j}$. The fluctuations in the phase are obtained from $\delta\alpha_{i,j}$ through $\delta\phi_{i,j} = \frac{\delta\alpha_{i,j}}{\bar{A}_{i,j}}$. 

Similar to the below threshold case, the correlations of the fluctuations in the signal and idler modes resulting from the pump drive are manifest in combined quadratures, 
\bea
\delta x_{\pm} &=&\frac{1}{\sqrt{2}}(\delta\alpha_{i}\pm\delta\alpha_{j}) \propto \delta\phi_{\pm}\label{Eq:phds}\\
\delta y_{\pm} &=& \frac{1}{\sqrt{2}}(\delta\beta_{i}\pm\delta\beta_{j}) =  \delta R_{\pm} \label{Eq:ampds}
\eea
where $\delta R_{\pm}$ are the amplitude sum and difference quadratures and $\delta\phi_{\pm}$ are the phase sum and difference quadratures. The spectrum of fluctuations of $\delta\alpha_{\pm}$ and $\delta\beta_{\pm}$ are given by Eq.[\ref{Eq:spectrum}].\\

In contrast to the dynamics below threshold, the fluctuations of the pump mode above threshold have an influence on the correlations generated between the membrane modes. On the one hand, the substrate fluctuations are smaller than those of the membrane modes by a term proportional to the respective mass ratios. Thus, one might assume that these tiny fluctuations should have a negligible influence on the membrane modes. However, the substrate mode fluctuations affect the membrane modes through terms that are proportional to the steady state amplitudes of the latter. These amplitudes are larger than the pump mode amplitude by the mass ratio of the substrate and membrane modes. These ratios cancel each other and lead to pump mode thermal fluctuations influencing the signal and idler modes through terms that are of the same order as the coupling between the signal and idler modes. Thus, the degree of amplitude difference squeezing is independent of the resonator to substrate mass ratio. 

Consistent with the experimental system under consideration \cite{patil2014}, we assume that the damping rate of the substrate excitation is 3-4 orders of magnitude larger than those of the membrane. In this regime, the pump fluctuations respond instantaneously to those of the membrane modes and can thus be adiabatically eliminated. 
We use this to simplify the analysis and ignore the time derivative of the pump fluctuations ($\delta \dot{A}_{S}$) in Eq.[\ref{Eq:fluc}]. Aside from this modification, we extract the fluctuations of the signal and idler modes as before.

\subsection{Matched losses and frequencies}
\begin{figure}[h]
\begin{center}
\includegraphics[width=0.45\textwidth]{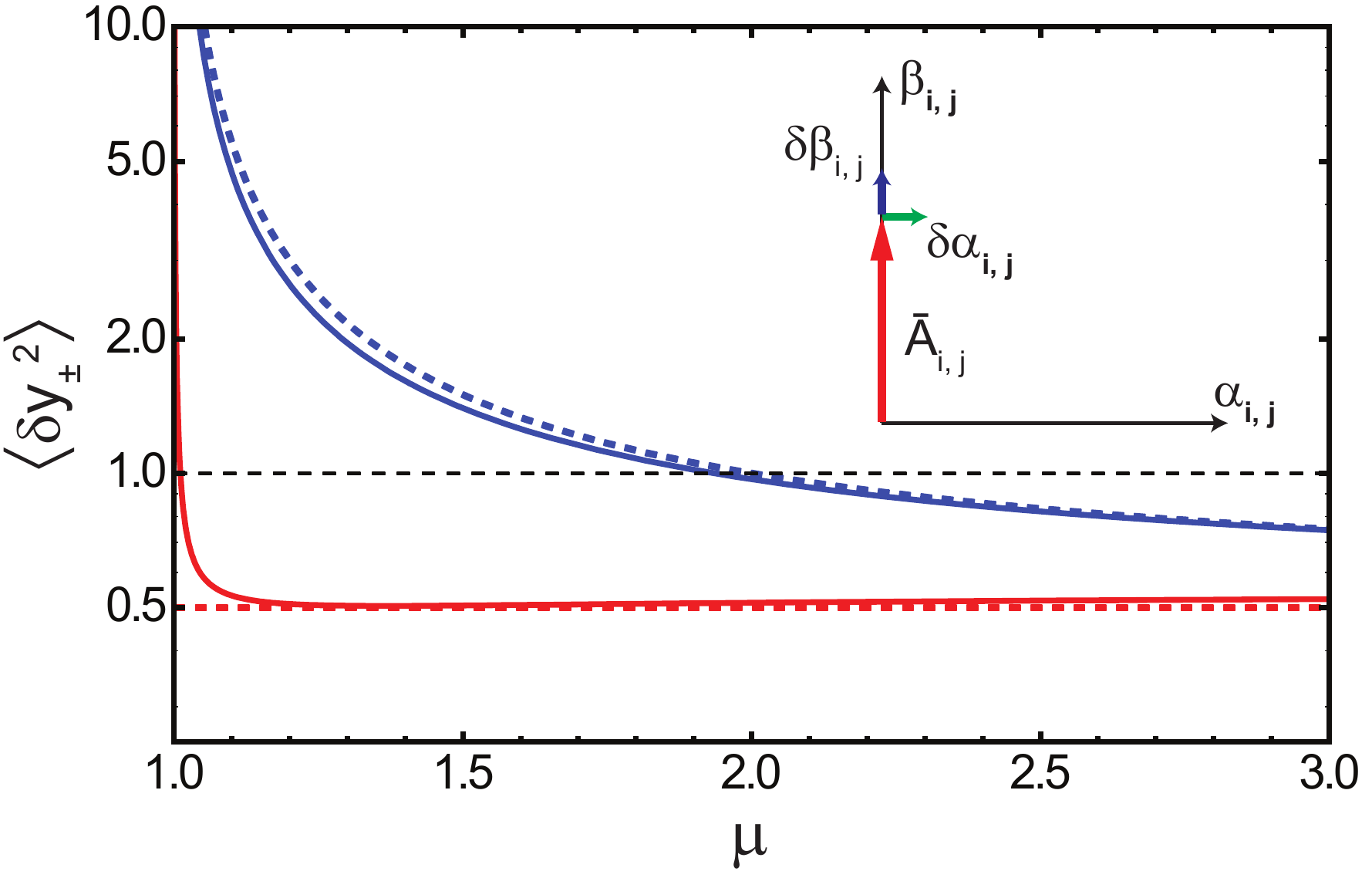}
\caption{Variance of the fluctuations in the difference $y_{-}$ (red) and sum $y_{+}$ (blue) quadratures {\em vs} the parametric drive $\mu$. The dashed lines are computed for $\delta_\gamma = \delta_\omega = 0$. The solid lines are computed for $\delta_\gamma=0.31, \delta_\omega=0.09$. The amplitude difference quadrature is squeezed for all values above threshold to a value of $\frac{1}{2}$. As can be seen, amplitude difference squeezing is extremely robust to experimental imperfections such as a frequency mismatch or loss asymmetry.\\
Inset: Schematic of the mean value of the membrane mode amplitude $\bar{A}_{i,j}$ and the fluctuations $\delta\alpha_{i,j}$ and $\delta\beta_{i,j}$. $\delta\vec{\beta}$ represent amplitude fluctuations while $\delta\vec{\alpha}$ are related to fluctuations of the phase.}
	\label{Fig4}
	\end{center}
\end{figure}

 We first consider the case where the damping rates and the frequencies are matched. In this limit, we obtain the following spectral densities for the collective quadratures normalized with respect to the thermal motion amplitude.
\bea
\mathbf{S}_{Y}(\omega) &=& \frac{1}{2\pi}\left(\begin{array}{cc} \frac{\gamma\mu}{\left(\gamma^{2}(-1+\mu)^{2}+\omega^2\right)} & 0\\ 0 & \frac{\gamma}{\left(\gamma^2+\omega^2\right)}\end{array}\right)  \\
\mathbf{S}_{X}(\omega) &=& \frac{1}{2\pi}\left(\begin{array}{cc} \frac{\gamma\mu}{\left(\gamma^{2}\mu^{2}+\omega^2\right)} & 0\\ 0 & \frac{\gamma}{\omega^2}\end{array}\right) 
\eea
We obtain the variances of the fluctuations by integrating the spectra. For the $Y$ quadrature, which relates to amplitude fluctuations, these evaluate to 
\bea
\sigma_{y_{+},y_{+}} &=&  \frac{\mu}{2\left(\mu-1\right)}\\
\sigma_{y_{-},y_{-}} &=&  \frac{1}{2}
\eea
These variances are plotted as a function of the parametric drive in Fig.[\ref{Fig4}].

Above threshold, the amplitude difference between the signal and idler modes is always half the thermal variance. This is the mechanical analogue of intensity difference squeezing seen in optical parametric oscillators. We find that while the individual amplitudes are sensitive to fluctuations of the pump mode, the amplitude difference is insensitive to fluctuations of the pump mode and the degree of squeezing is independent of the pump drive. On the other hand, the variance of the amplitude sum diverges as $\mu \rightarrow 1^{+}$ and decreases with increasing drive, approaching half the thermal variance as $\mu \rightarrow \infty$.

The other notable feature of above threshold dynamics is that the phase difference between the signal and idler modes is unspecified and hence free to fluctuate. The fluctuation in the phase difference is given by 
\bea
S_{\phi_-,\phi_-}(\omega) &=& \left(\frac{x^{2}_{th}}{A(\mu)^{2}}\right)S_{x_{-},x_{-}}(\omega)\\
&=& \left(\frac{x^{2}_{th}}{A(\mu)^{2}}\right) \left(\frac{\gamma}{2\pi\omega^2}\right)
\eea
where $A(\mu)$ is the amplitude of the membrane modes (identical for the case of matched loss rates), $x_{th}^{2} = \frac{k_B T}{m \omega^2}$ is the thermal variance. The integral of the fluctuation spectrum diverges as $\omega^{-2}$, indicating that the difference phase undergoes diffusion. The time scale $\tau$ for this diffusion can be estimated by calculating the variance while imposing a low frequency cut off ($\frac{2\pi}{\tau}$) to the integral of the spectral density, i.e.
\bea
\langle \delta\phi_{-}^{2}\rangle &=& 2\int^{\infty}_{\frac{2\pi}{\tau}}{S_{\phi_-,\phi_-}(\omega)}d\omega\\
&=& \left(\frac{x^{2}_{th}}{A(\mu)^{2}}\right) \frac{\gamma \tau}{2\pi^{2}}
\eea 
If the parametric nonlinearity were used to actuate the membrane modes to an amplitude of $10^3 \times (x^2_{th})^{1/2}$, a phase fluctuation of 1 mrad would require a measurement duration of 10 ringdown periods or around 100 seconds for the modes considered in \cite{patil2014}. Thus, while $S_{\phi_-,\phi_-}(\omega)$ diverges, it does not necessarily lead to large fluctuations of the difference phase over experimental time scales. 

\subsection{Effect of mismatched frequencies and loss rates}
For the case of non-zero loss asymmetry and frequency mismatch ($\delta_\gamma, \delta_{\omega} \neq 0$), the fluctuations of the amplitude difference are no longer decoupled from fluctuations of the amplitude sum. These fluctuations are obtained as in the previous section. The calculation, while straightforward, is laborious and we only summarize the results below. 

The variances of the amplified and squeezed sum and difference quadratures are shown in Fig.[6] for $\delta_\gamma = 0.31$ and  $\delta_{\omega} = 0.09$.  In this case, the coupling between amplified and squeezed quadratures leads to a divergence in the squeezed quadrature as $\mu \rightarrow 1^{+}$. Unlike in the case below threshold,  the fluctuations for the case of matched asymmetries ($\delta_\gamma = \delta_{\omega} \neq 0$) are not the same as for the case of $\delta_\gamma=\delta_{\omega} = 0$. Importantly, we note that for pump actuation significantly above threshold, the degree of squeezing is impervious to experimental imperfections such as a loss asymmetry. 

\section{Crossover of correlations at the instability threshold}

Finally, we discuss the dynamics of the two-mode nonlinearity in the vicinity of the parametric instability. The crossover regime is most conveniently portrayed by evaluating the variance of $y_\pm$ quadratures in the regimes below and above threshold. In the former regime, these quadratures represent the two-mode correlations arising from phase-sensitive parametric amplification. In the latter regime, these correspond to the sum and difference amplitude fluctuations of the two membrane modes. These are shown in Fig.[7]. In the general case of mismatched dissipation rates ($\delta_\gamma \neq 0$), both these quadratures exhibit diverging variances at the instability. 

\begin{figure}[h]
\begin{center}
\includegraphics[width=0.45\textwidth]{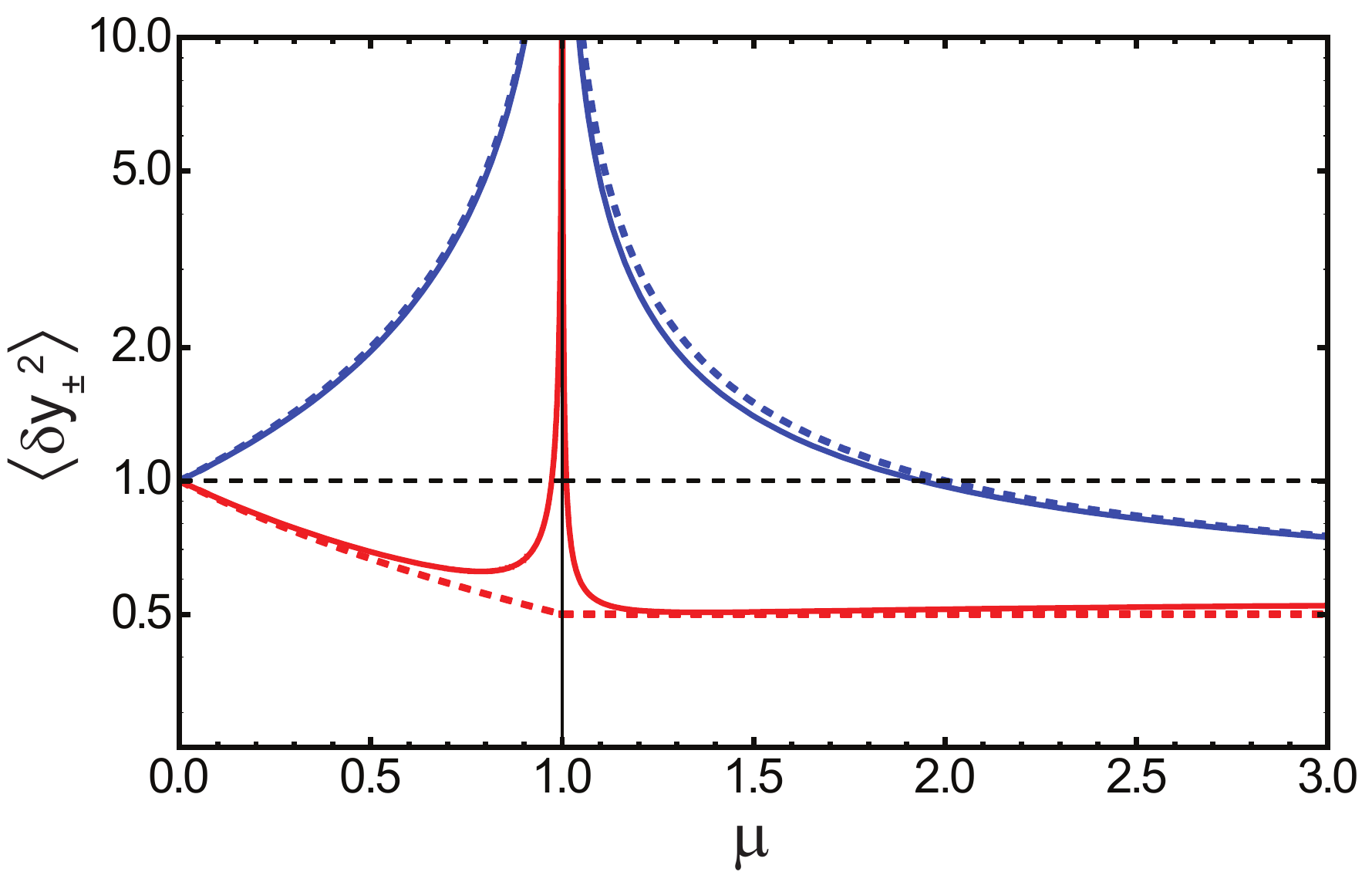}
\caption{Two-mode correlations in the vicinity of the threshold for parametric instability. Variance of the fluctuations of the normalized difference ($y_{-}$) and sum ($y_{+}$) quadratures above and below threshold {\em vs} the normalized parametric drive $\mu$. The dashed lines represent the variances for $\delta_\gamma = \delta_\omega = 0$. The solid lines represent the variances for $\delta_\gamma = 0.31, \delta_\omega = 0.09$. The dashed horizontal line represents the thermomechanical variance given by $\frac{k_B T}{m \omega^2}$.}
	\label{Fig5}
	\end{center}
\end{figure}

As for the divergent phase diffusion discussed earlier, finite time effects need to be considered to interpret the divergent steady-state variances depicted in Fig.[7]. For the low frequency, ultrahigh quality factor resonators considered in this work, the divergent response time in the vicinity of the instability threshold can result in inordinately long thermalization times ($\sim 10^4 - 10^5$ seconds). For typical measurement durations ($\sim 100$ seconds), the measured squeezing spectra can deviate appreciably from the spectra computed in steady state. Expectedly, these deviations are most significant for parametric actuation around the instability threshold ($\mu \sim 1.0$). 

\begin{figure}[h]
\begin{center}
\includegraphics[width=0.45\textwidth]{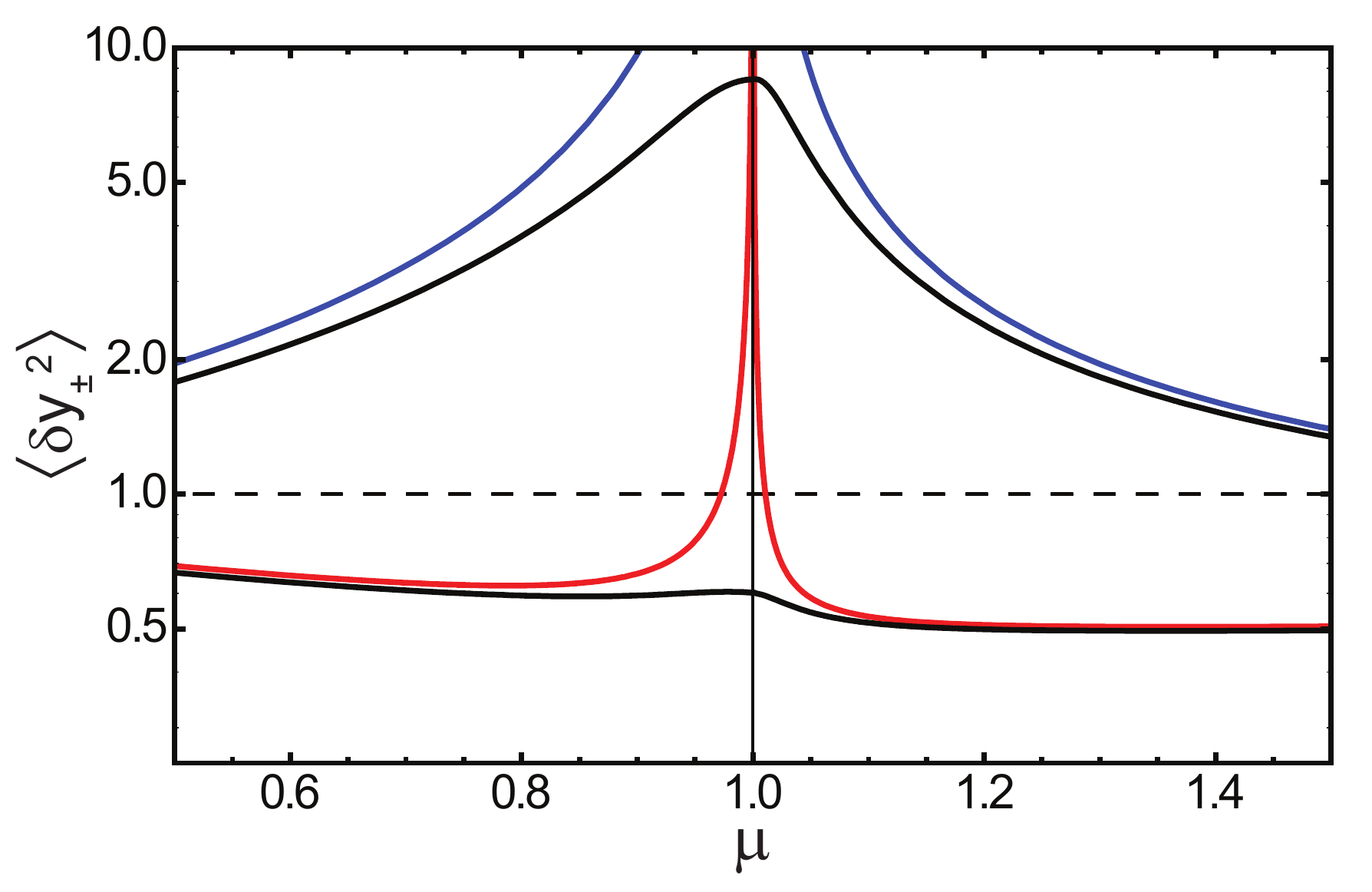}
\caption{Corrections to squeezing spectra due to finite measurement duration. Normalized variance of amplified ($y_{+}$) and squeezed ($y_{-}$) quadratures for a measurement time of $300$ seconds ($\sim$ 100 ring down periods) (black solid lines).  Also shown for comparison are the corresponding amplified (blue) and squeezed (red) quadratures in steady state. These are computed for $\delta_\gamma = 0.31$, $\delta_\omega = 0.09$ ($\gamma \sim 2\pi\times 100$ mHz) and correspond to the experimental parameters in \cite{patil2014}. The divergence at the instability threshold is attenuated for finite measurement times due to the divergent response times.}
\label{Fig_finitetime}
\end{center}
\end{figure}

The variances measured over a finite time $\tau_m$ are extracted by truncating the integral of the relevant spectral densities by the time of measurement, i.e.
\be
\boldsymbol{\sigma}_{\alpha,\beta} = 2\int_{\frac{2\pi}{\tau_{m}}}^{\infty}\mathbf{S}_{\alpha,\beta}(\omega)d\omega \label{wkF}
\ee 
These variances, computed for the parameters in \cite{patil2014}, result in the solid black curves shown in Fig.[\ref{Fig_finitetime}]. We see that the singularities in the amplified and squeezed quadratures seen in the steady state variances (solid blue and red lines) are washed out at finite measurement durations. 

The regime around $\mu = 1$ is of interest since the system exhibits mechanical bistability and a hysteretic response due to the divergent mechanical susceptibility and diverging response time. This regime offers a clean, mesoscopic and mechanical realization of a second-order phase transition \cite{gerry1988, mertens1993, kinsler1995, dechoum2004} and out-of-equilibrium quantum dynamics on experimentally accessible timescales. Furthermore, quantum tunneling between bistable mechanical states has been discussed in the context of a Duffing nonlinearity as a means of accessing a quantum-to-classical transition \cite{katz2008}. In the system considered here and in \cite{patil2014}, the presence of a strong two-mode nonlinearity and compatibility with optomechanical cooling together imply that similar effects can be accessed even in the regime of low phonon number and motion on the scale of mechanical zero-point fluctuations. The nature of this nonequilibrium phase transition and the response to critical fluctuations as the phonon occupancy is gradually reduced by optomechanical cooling will be described elsewhere. 

We also note the close correspondence between the squeezing spectra arising from the two-mode nonlinearity and the properties of the reservoir to which the resonator is coupled. In particular, while we have considered the squeezing spectra in the presence of a Markovian (thermal) reservoir in this work, it is also known that intrinsic defects or two-level systems (TLS) in amorphous resonators such as SiN can couple to mechanical motion \cite{ramos2013}. Furthermore, a reservoir of such TLS can acquire non-Markovian properties in certain regimes. In addition to studying the effect of such non-Markovian fluctuations on the second order phase transition and the two-mode correlations near the instability, it is an intriguing prospect to use the exquisite sensitivity of this two-mode nonlinearity as an amplifier of such reservoir interactions to shed light on the intrinsic material properties of the membrane resonator. 

\section{Summary\label{sec:7}}
In summary, we describe a phononic nondegenerate parametric amplifier that is realized in a membrane resonator through a substrate-mediated nonlinearity. Motivated by recent work \cite{patil2014}, we discuss the creation of multimode phononic correlations arising from this parametric interaction and compute the squeezing spectra in the presence of thermomechanical noise. We address various points of experimental relevance including the presence of frequency mismatches and asymmetric dissipation rates between the resonator modes. We find the robust presence of  two-mode phononic correlations for a wide range of experimental parameters.

Below the threshold for parametric instability, this system exhibits two-mode noise squeezing of collective quadratures composed from the individual resonator modes. This regime is conducive to back-action evading (BAE) schemes for quantum-enhanced metrology. Above threshold, the system exhibits amplitude difference squeezing due to the correlated production of down-converted phonons in both resonator modes. This regime is promising for nonlinear measurement schemes capable of super-Heisenberg sensitivities. 

In the crossover regime between these two limits, the response of the system near the parametric instability threshold is characterized by a divergent mechanical susceptibility and a diverging response time. In this regime, the system exhibits mechanical bistability, a hysteretic response and critical slowing down. This regime is of interest as it offers a mechanical realization of a second order phase transition for the investigation of nonequilibrium critical dynamics, the quantum-to-classical transition and the influence of a non-Markovian reservoir on such critical phenomena. These results will be discussed elsewhere. Due to the combined presence of low intrinsic dissipation and optomechanical compatibility, such nonlinear mechanical systems are promising for extending these studies from the classical realm deep into the quantum regime. 

This work was supported by the DARPA QuASAR program through a grant from the ARO, the ARO MURI on non-equilibrium Many-body Dynamics (63834-PH-MUR), an NSF INSPIRE award and the Cornell Center for Materials Research with funding from the NSF MRSEC program (DMR-1120296). We acknowledge valuable discussions with H. F. H. Cheung. M. V. acknowledges support from the Alfred P. Sloan Foundation. 

\bibliography{SiNbib}

\section*{Appendix I: Calculation of fluctuation spectra in the presence of thermal noise\label{sec:3}}
For both the below and the above threshold regimes, we obtain the correlations that develop between the resonator modes in the presence of the pump drive, by analyzing the coupled equations for the resonator modes under the influence of a classical actuation of the pump/substrate mode along with thermomechanical Langevin noise forces acting on the membrane and substrate modes.

We distinguish between the mean displacement and the fluctuations by writing $x_{i,j} = (\bar{A}_{i,j} + \delta A_{i,j}) e^{-i \omega_{i,j}t}$ where $\langle \delta A_{i,j} \rangle = 0$.
The coupled equations for the fluctuations can be written as,
\bea
&2 \left( \begin{array}{c}
	\delta \dot{A}_i \\ \delta \dot{A}_j \\ \delta \dot{A}_S \end{array} \right) = \left( \begin{array}{ccc} -\gamma_i & 0 & i \kappa_{i} \bar{A}_j^* \\
	0 & -\gamma_j & i \kappa_{j} \bar{A}_i^* \\ i \kappa_{S} \bar{A}_j & i\kappa_{S} \bar{A}_i & -\gamma_S \end{array}\right) \left( \begin{array}{c}
	\delta A_i \\ \delta A_j \\ \delta A_S \end{array} \right) \nonumber &\\
&+  \left( \begin{array}{ccc} 0 & i\kappa_{i} \bar{A}_S & 0 \\
	i \kappa_{j} \bar{A}_S & 0 & 0 \\ 0 & 0 & 0 \end{array}\right) \left( \begin{array}{c}
	\delta A^*_i \\ \delta A^*_j \\ \delta A^*_S \end{array} \right) + i \left( \begin{array}{c}
	\gamma_i \chi_i F_i \\ \gamma_j \chi_j F_j \\ \gamma_S \chi_S F_S \end{array} \right)\label{Eq:fluc} \nonumber
\eea
where we have defined coupling parameters, 
\begin{eqnarray}
\kappa_{k}&=& \frac{g\gamma_{k}\chi_{k}}{2} = \frac{g}{2m_{k}\omega_{k}} ;\quad k\in[i,j,S] \label{kappa}
\eea
for notational simplicity. The thermomechanical noise forces are assumed to be white noise correlated and obey,
\begin{eqnarray}
\langle F_i(t) \rangle = \langle F_i (t) F_j (t') \rangle = 0, \\
\langle F_i (t) F^{*}_j (t+\tau) \rangle = 8 \gamma_i m_i k_B T \delta_{ij} \delta(\tau)
\end{eqnarray}
We decompose the complex displacements into real quadratures according to $\delta {\bf A} = \delta{\vec \alpha} + i \delta{\vec\beta}$ and 
where $\delta{\bf A} = (\delta A_i, \delta A_j, \delta A_S)^T$. Correspondingly the noise, $\mathbf{v} = i (\gamma_{i}\chi_{i}F_{i},\gamma_{j}\chi_{j}F_{j},\gamma_{S}\chi_{S}F_{S})^{T}$is also decomposed into real and imaginary parts, $\mathbf{v} = \mathbf{v}_{\alpha} + i\mathbf{v}_{\beta}$ . Expressing Eq.[\ref{Eq:fluc}] in terms of these quantities gives,  
\begin{eqnarray}
\delta \dot{\vec \alpha} &=& {\bf M}_\alpha \delta {\vec \alpha} + {\bf v}_\alpha \label{Eq:alpha} \\
\delta \dot{\vec \beta} &=& {\bf M}_\beta \delta {\vec \beta} + {\bf v}_\beta \label{Eq:beta}
\end{eqnarray}
For the general case, valid both above and below threshold, 
\begin{equation}
{\bf M}_{\alpha/\beta} = \frac{1}{2} \left( \begin{array}{ccc} -\gamma_i & \mp\kappa_{i}|\bar{A}_S| & \kappa_{i}|\bar{A}_j| \\
\mp \kappa_{j}|\bar{A}_S| & -\gamma_j &  \kappa_{j}|\bar{A}_i| \\
-\kappa_{S}|\bar{A}_j| & -\kappa_{S}|\bar{A}_i| & -\gamma_S \end{array}\right)\label{Mgen}
\end{equation}
and the elements of ${\bf v}_{\alpha,\beta}$ satisfy $\langle v_i \rangle = 0$, $\langle v_k (t) v_l (t + \tau) \rangle = \frac{\gamma_l k_B T}{m_l \omega_l^2}\delta_{kl} \delta (\tau)$. In writing Eq.[\ref{Mgen}], we have made a choice for the pump drive phase ($\phi_{S} = 0$) and the resonator mode detection phases ($\phi_{i,j} = 0$).  In general, these phases can also be chosen such that there is a coupling between the $\delta\vec{\alpha}$ and $\delta\vec{\beta}$ quadratures.  When the pump drive phase is not fixed, but evolving in time, for instance with the pump drive being detuned, this coupling is physical and cannot be made to vanish through a suitable choice of detection phases. 

The noise spectral density in the steady state is obtained by taking the expectation value after Fourier transforming and inverting Eqs.[38,39], and are given by the matrix equation,
\begin{equation}
{\bf S}_{\alpha/\beta}(\omega) = \frac{1}{2 \pi} ({\bf M}_{\alpha/\beta} + i \omega {\bf I})^{-1} {\bf D} ({\bf M}^T_{\alpha/\beta} - i \omega {\bf I})^{-1} \label{Eq:spectrum}
\end{equation}
where ${\bf I}$ is the identity and 
\begin{equation}
{\bf D} = \langle {\bf v} {\bf v}^T \rangle = k_B T \left( \begin{array}{ccc} \frac{\gamma_i}{m_i \omega_i^2} & 0 & 0 \\ 0 & \frac{\gamma_j}{m_j \omega_j^2} & 0 \\ 0 & 0 & \frac{\gamma_S}{M_{S} \omega_S^2} \end{array}\right)
\end{equation}
is a matrix characterizing the diffusion due to thermal forces. The variances in steady state can be obtained from the spectrum using the Wiener-Khintchine theorem, by integrating the fluctuations over frequency, i.e.
\be
\boldsymbol{\sigma}_{\alpha/\beta} = \int_{-\infty}^{\infty}\mathbf{S}_{\alpha/\beta}(\omega)d\omega \label{wk}
\ee 

\section*{Appendix II : Calculation of fluctuation spectra for finite pump detuning}
For finite pump detuning below threshold, the actuating force is a slowly varying function of time, i.e. $\tilde{F}_S = |\tilde{F}_S|e^{-i \Delta t}$ where the drive frequency $\omega_d = \omega_S + \Delta$. This results in a pump amplitude given by $\bar{A}_S = i |\bar{A}_S| e^{-i \Delta t}$. 

Linearizing about the steady state amplitude, i.e $A_{k} = \bar{A}_{k} + \delta A_{k}(t)$ where $k \in [i,j,S]$, with $\bar{A}_{i,j} = 0$, and defining the vectors $\delta\vec{A} = (\delta A_{i}, \delta A_{j})^{T}$ and $\delta\vec{v} = (v_{i}, v_{j})^{T}$, the relevant equations of motion for the fluctuations of the resonator modes reduce to,
\begin{eqnarray}
2\dot{\delta\vec{A}}
&=& -\begin{pmatrix}
\gamma_{i} && 0 \\
0 && \gamma_{j} \
\end{pmatrix}\delta\vec{A}\nonumber\\
&-&
\begin{pmatrix}
0 && \kappa_{i}|\bar{A}_{S}|e^{-i\Delta t}\\
\kappa_{j}|\bar{A}_{S}|e^{-i\Delta t} && 0 \
\end{pmatrix}
\delta\vec{A}^{*}+2\vec{v}\label{Eq:Avec}
\end{eqnarray}
By going to a frame rotating at $\frac{\Delta}{2}$, we rewrite $\delta\vec{A} = \delta\vec{B}e^{\frac{i\Delta t}{2}}$
in terms of which Eq.[\ref{Eq:Avec}] becomes,
\bea
2\dot{\delta\vec{B}}
&=& -\begin{pmatrix}
\gamma_{i} && 0 \\
0 && \gamma_{j} \
\end{pmatrix}\delta\vec{B} -i\Delta\delta\vec{B} \nonumber\\
&-&
\begin{pmatrix}
0 && \kappa_{i}|\bar{A}_{S}|\\
\kappa_{j}|\bar{A}_{S}| && 0 
\end{pmatrix}
\delta\vec{B}^{*}+ 2\vec{v}e^{-\frac{\Delta t}{2}}\label{Eq:Bvec}
\end{eqnarray}
where $\delta\vec{B}$ are the complex amplitudes of motion, measured at frequencies that are detuned from the individual mechanical modes $\omega_{i,j}$ by $\frac{\Delta}{2}$. We rewrite the complex amplitudes in terms of the real quadratures $\alpha_{i,j},\beta_{i,j}$ and decompose the noise term into real and imaginary parts, i.e. $\delta\vec{B} = \delta\vec{\alpha}+i\delta\vec{\beta}$ and $\vec{v} =  \vec{v}_{\alpha} + i\vec{v}_{\beta}$, in terms of which the equations of motion become,
\begin{eqnarray}
\delta \dot{\vec \alpha} &=& {\bf M}_\alpha \delta {\vec \alpha} -\frac{\Delta}{2}\delta\vec{\beta} + \vec{v}_\alpha \label{Eq:alpha}\\
\delta \dot{\vec \beta} &=& {\bf M}_\beta \delta {\vec \beta} +\frac{\Delta}{2}\delta\vec{\alpha} +\vec{v}_\beta \label{Eq:beta}
\end{eqnarray}
where 
\bea
{\bf M}_{\alpha/\beta} &=& \frac{1}{2}\begin{pmatrix}
-\gamma_{i} && \mp\kappa_{i}|\bar{A}_{S}|\\
\mp\kappa_{j}|\bar{A}_{S}| && -\gamma_{j}
\end{pmatrix}
\eea
and the elements of $\vec{v}_{\alpha,\beta}$ satisfy $\langle v_{k,\eta} \rangle \quad \eta \in [\alpha, \beta]; k \in [i,j]$ and $\langle v_{k,\eta} (t) v_{l,\eta'} (t + \tau) \rangle = \frac{\gamma_l k_B T}{m_l \omega_l^2}\delta_{kl}\delta_{\eta,\eta'} \delta (\tau)$.

The coupling between the $\delta\vec{\alpha}$ and $\delta\vec{\beta}$ quadratures of the individual oscillators resulting from the detuned drive is apparent in the above equations. The steady state correlations between these quadratures can be obtained by forming the following 4 dimensional vectors; $\mathbf{Z} = (\delta\alpha_{i},\delta\alpha_{j},\delta\beta_{i},\delta\beta_{j})^{T} = (\delta\vec{\alpha},\delta\vec{\beta})^{T}$ and $\mathbf{v} = (\vec{v}_{\alpha},\vec{v}_{\beta})^{T}$, in terms of which the equations of motion become,
\bea
\dot{\mathbf{Z}} &=& \mathbf{M}\mathbf{Z} + \mathbf{v} \label{Eq:X}\\
\mathbf{M} &=& \begin{pmatrix}
\mathbf{M}_{\alpha} && -\frac{\Delta}{2}\mathbf{I}\\
\frac{\Delta}{2}\mathbf{I} && \mathbf{M}_{\beta}
\end{pmatrix} 
\eea
and $\mathbf{I}$ is the $2\times2$ identity matrix.

The noise spectral densities are obtained by solving Eq.[\ref{Eq:X}] in fourier space, as before. The spectrum in the steady state is,
\bea
{\bf S}(\omega) &=& \left\langle\mathbf{Z}(\omega)\mathbf{Z}(\omega)^{\dagger}\right\rangle \nonumber\\
&=& \frac{1}{2 \pi} ({\bf M} + i \omega {\bf I})^{-1} {\bf D} ({\bf M}^T - i \omega {\bf I})^{-1}
\eea
where ${\bf I}$ is the identity and 
\begin{equation}
{\bf D} = \langle {\bf v} {\bf v}^T \rangle = k_B T \left( \begin{array}{cccc} \frac{\gamma_i}{m_i \omega_i^2} & 0 & 0 & 0 \\ 0 & \frac{\gamma_j}{m_j \omega_j^2} & 0 & 0\\ 0 & 0 & \frac{\gamma_i}{m_{i} \omega_i^2} & 0 \\ 0 & 0 & 0 & \frac{\gamma_j}{m_{j} \omega_j^2}\end{array}\right)
\end{equation}

We construct composite quadratures  $x_{\pm} = (\alpha_{i}\pm\alpha_{j})/\sqrt{2}$, $y_{\pm} = (\beta_{i}\pm\beta_{j})/\sqrt{2}$ as before and represent the fluctuations in these quadratures by the column matrix, $\mathbf{Z}_{c} = (\delta x_{+},\delta x_{-},\delta y_{+},\delta y_{-})^{T}$, which is related to $\mathbf{Z}$ by 
\bea 
\mathbf{Z}_{c} = \mathbf{R}\mathbf{Z}; \quad\mathbf{R} = \frac{1}{\sqrt{2}}\left( \begin{array}{cccc} 1 & 1 & 0 & 0 \\ 1 & -1 & 0 & 0 \\ 0 & 0 & 1 & 1 \\ 0 & 0 & 1 & -1 \end{array}\right)
\eea
The correlations of the composite quadratures are therefore given by,
\bea
\mathbf{S}_{c}(\omega) = \left\langle \mathbf{Z}_{c}(\omega)\mathbf{Z}_{c}(\omega)^{\dagger} \right\rangle =  \mathbf{R}\mathbf{S}\mathbf{R}^{T}
\eea
We consider the case where the frequencies of the resonator modes are identical and the losses are symmetric ($\delta_\gamma = \delta_\omega = 0$). For this case, the diffusion matrix $\mathbf{D} = \frac{k_B T \gamma}{m\omega^{2}}\delta_{ij}$. 

The correlations between the composite quadratures in this case are given by,
\be
\mathbf{S}_{c}(\omega) = \left( \begin{array}{cccc} S_{x_{+},x_{+}} & 0 & S_{x_{+},y_{+}} & 0 \\ 0 & S_{x_{-},x_{-}} & 0 & S_{x_{-},y_{-}}\\ S_{y_{+},x_{+}} & 0 & S_{y_{+},y_{+}}  & 0 \\ 0 & S_{y_{-},x_{-}} & 0 & S_{y_{-},y_{-}} \end{array}\right)\label{Eq:Spm}
\ee
where the non zero correlations are as indicated above. There are no correlations between ($x_{+}$, $x_{-}$) and ($y_{+}$, $y_{-}$), given our choice of detection phases and the fact that we consider the case $\delta_\gamma = \delta_{\omega} = 0$. The correlations in Eq.[\ref{Eq:Spm}] evaluate to,
\bea
S_{x_{\pm},x_{\pm}}(\omega) &=& S_{y_{\mp},y_{\mp}}(\omega) \nonumber \\
&=& \frac{k_B T \gamma}{m \omega^2}\, \frac{2 \left(\Delta ^2+\gamma ^2 (1\mp\mu)^2+4\omega^2\right)}{\pi\left(4\omega^2+\lambda_{+}^{2}\right)\left(4\omega^2+\lambda_{-}^2\right)}\\
S_{x_{+},y_{+}}(\omega) &=& S_{y_{+},x_{+}}(\omega)\nonumber \\
&=& \frac{k_B T \gamma}{m \omega^2}\, \frac{4 \Delta\left(\gamma\mu+2i\omega\right)}{\pi\left(4\omega^2+\lambda_{+}^{2}\right)\left(4\omega^2+\lambda_{-}^2\right)}\\
&=& S_{x_{-},y_{-}}(-\omega) = S_{y_{-},x_{-}}(-\omega)\nonumber 
\eea
where $\lambda^{2}_{\pm} =\gamma ^2\left(1+\mu ^2\right)-\Delta ^2 \pm 2\gamma\sqrt{\gamma ^2\mu ^2-\Delta ^2}$. The variances in the steady state obtained using the Weiner-Khintchine theorem are,
\bea
\sigma_{x_{\pm},x_{\pm}} &=& \left(\frac{k_{B}T\gamma}{m\omega^{2}}\right)\Bigg[\frac{\left(\Delta ^2+\gamma ^2 (1\mp\mu)^2-\lambda_{-}^2\right)}{\lambda_{+}\lambda_{-}\left(\lambda_{+}+\lambda_{-}\right)} + \frac{1}{\lambda_{+}}\Bigg]\nonumber\\
&=& \sigma_{y_{\mp},y_{\mp}}
\eea
Unlike the case of zero detuning, we obtain non-zero steady state correlations between the  $x$ and $y$ quadratures,
\bea
\sigma_{x_{\pm},y_{\pm}} &=& \left(\frac{k_{B}T\gamma}{m\omega^{2}}\right) \frac{2\Delta\gamma\mu}{\lambda_{+}\lambda_{-}\left(\lambda_{+}+\lambda_{-}\right)} = \sigma_{y_{\pm},x_{\pm}}
\eea
The correlations between the $x_{\pm}$ and $y_{\pm}$ are now proportional to the drive detuning. 

\end{document}